\documentclass[10pt,conference]{IEEEtran}
\usepackage{makecell}
\usepackage{hyperref}

\IEEEoverridecommandlockouts
\usepackage{amssymb,amsmath,amsthm}

\makeatletter
\newcommand{\biggg}[1]{{\hbox{$\left#1\vbox to 20.5pt{}\right.\n@space$}}}
\newcommand{\Biggg}[1]{{\hbox{$\left#1\vbox to 23.5pt{}\right.\n@space$}}}
\newcommand{\bigggg}[1]{{\hbox{$\left#1\vbox to 26.5pt{}\right.\n@space$}}}
\newcommand{\Bigggg}[1]{{\hbox{$\left#1\vbox to 29.5pt{}\right.\n@space$}}}
\newcommand{\biggggg}[1]{{\hbox{$\left#1\vbox to 32.5pt{}\right.\n@space$}}}
\newcommand{\Biggggg}[1]{{\hbox{$\left#1\vbox to 35.5pt{}\right.\n@space$}}}
\newcommand{\bigggggg}[1]{{\hbox{$\left#1\vbox to 38.5pt{}\right.\n@space$}}}
\newcommand{\Bigggggg}[1]{{\hbox{$\left#1\vbox to 41.5pt{}\right.\n@space$}}}
\makeatother

\usepackage{bm,cite,algorithm,algorithmic,float,amsmath,amssymb}
\usepackage[dvips]{graphicx}
\usepackage{subfigure,amsfonts}
\usepackage{mdwmath}
\usepackage{mdwtab}
\usepackage{xcolor}
\usepackage{cool}
\usepackage{makeidx,enumitem}
\usepackage{balance}
\floatname{algorithm}{Algorithm}

\makeatletter
\renewcommand\paragraph{\@startsection{paragraph}{4}{\z@}%
            {-2.5ex\@plus -1ex \@minus -.25ex}%
            {1.25ex \@plus .25ex}%
            {\normalfont\normalsize\itshape}}
\makeatother
\usepackage{epstopdf}

\begin{document}
\title{A Machine Learning Solution for Beam
Tracking in mmWave Systems \thanks{This work was supported financially in part by the National Science Foundation (NSF CIF-1618078).}
 \thanks{The authors are at the Ming Hsieh Department of Electrical Engineering, University of Southern California, 
Los Angeles, CA 90089, USA
(email: burghal,nabbasi,molisch@usc.edu).}}
\author{
\IEEEauthorblockN{Daoud Burghal, \emph{Student Member, IEEE,} Naveed A. Abbasi, \emph{Member, IEEE,} and Andreas F. Molisch, \emph{Fellow, IEEE}}}
\maketitle

\begin{abstract}
Utilizing millimeter-wave (mmWave) frequencies for wireless communication in \emph{mobile} systems is challenging since it requires continuous tracking of the beam direction. Recently, beam tracking techniques based on channel sparsity and/or Kalman filter-based techniques were proposed where the solutions use assumptions regarding the environment and device mobility that may not hold in practical scenarios. In this paper, we explore a machine learning-based approach to track the angle of arrival (AoA) for specific paths in realistic scenarios. In particular, we use a recurrent neural network (R-NN) structure with a modified cost function to track the AoA. We propose methods to train the network in sequential data, and study the performance of our proposed solution in comparison to an extended Kalman filter based solution in a realistic mmWave scenario based on  stochastic channel model from the QuaDRiGa framework. Results show that our proposed solution outperforms an extended Kalman filter-based method by reducing the AoA outage probability, and thus reducing the need for frequent beam search. 
\end{abstract}
\section{Introduction}
Next-generation wireless standards aim to utilize millimeter-wave (mmWave) frequencies as one of the key avenues to meet the ever-increasing demands for bandwidth. Due to channel impairments at high frequencies such as high path loss, diffraction loss etc. \cite{Molisch2012Wireless}, using multi-antenna systems is necessary. This in turn requires proper beam management techniques to harness the advantages of mmWave band. More specifically, beam management techniques can be grouped in two main categories: beam-search and beam-steering, where the former refers to the procedures needed during the initial communication link establishment or after a link failure, while the latter refers to the techniques used to steer a beam to different directions. Beam management usually consumes considerable overhead as both ends may need to exchange pilot signals to steer the beams in the right direction.

The overhead problem is exacerbated in mobile systems where the optimal beam direction changes over time due to the mobility of the user equipment (UE). Several recent works propose techniques to track the beam directions, \cite{palacios2017tracking,jayaprakasam2017robust,va2016beam,he2014millimeter,gao2016beampattern,ramasamy2012compressive,zhang2016tracking,duan2015aod,larew2019adaptive}. In \cite{duan2015aod} and \cite{ramasamy2012compressive}, the authors make use of channel sparsity in mmWave band to simplify the beam tracking problem. In \cite{jayaprakasam2017robust,va2016beam,zhang2016tracking,larew2019adaptive}, the authors use Kalman filter (KF)-based methods to track the beam directions. However, these methods have several issues, such as the intensive calculations required by sparse optimization problems that might not be suitable for implementation especially at the UE side, or the simplified channel models that are used in KF-based methods. The authors of \cite{palacios2017tracking,he2014millimeter,gao2016beampattern} use prior beam direction to narrow down the search space for efficient tracking; however, these methods do not fully capture the true channel dynamics. 

In this paper, we explore a machine learning (ML) approach for the beam tracking problem. In particular, we use a recurrent neural network (R-NN) to track the angle of arrival (AoA) at the UE side when only the channel coefficient and previous estimates of AoAs are available. The main advantages of using ML over other techniques are twofold: (i) based on the Universal Approximation property of NNs, we anticipate that the proposed solution can capture the channel and environment dynamics well, and (ii) low complexity of mathematical operations. For instance, the complex mathematical operations needed by "ML training", can be done offline at the base-station (BS) side, and the trained solution can be relayed to UEs as required. Once the training has been completed, ML solutions typically require basic mathematical operations that can be performed easily in real-time. 

Applying ML to wireless systems has gained considerable attention lately, where applications span different layers of communication systems \cite{zamanipour2019survey,o2017introduction, bogale2018machine, mao2018deep}, such as scheduling and resource allocation \cite{cao2018machine,burghal2018band}, detection, decoding\cite{nachmani2016learning}, and estimation \cite{zamanipour2019survey}. There have been recent efforts to apply ML to estimate the AoA. In \cite{hu2019low}, the authors use Deep NNs to provide a candidate set of angle estimation that can be used as an input to a maximum likelihood estimator for optimal angle search and selection. The authors of \cite{navabi2018predicting} use observable features at the BS side to predict the AoA at the UE side. In \cite{wu2019deep}, the authors take sparsity of the incident signals into consideration to propose a ML solution that solves the transformation from array outputs to AoA spectrum. Moreover, in \cite{liu2018direction}, the authors propose a ML solution based on autoencoders and followed NN layers for robust AoA estimation. 

In contrast to prior work, we tackle the problem of AoA \em{tracking} for \em{mobile} user over a path. In such cases, the UE parameters are in the form of sequences that evolve over time,  which calls for the selection of suitable recurrent ML solutions. We, therefore, use the long short term memory (LSTM) network architecture for the AoA tracking problem. Moreover, due to the periodicity of the angles, we propose a cosine-based loss function. We study the performance of the network in a stochastic channel model that we generate using the QuaDRiGa framework \cite{jaeckel2014quadriga}, and we also compare the performance of the ML solution against the KF-based solution proposed in \cite{va2016beam}. 

The structure of the paper is as follows. In Sec. \ref{sec:prob}, we provide the problem setup and summarize the KF based solution of \cite{va2016beam}. We introduce the ML solution and the cost function in \ref{sec:ML}. In Sec. \ref{sec:Data}, we discuss the method used to generate the data and train the ML solution. We analyze the performance of our solution in Sec \ref{sec:Results}, and finally, provide the concluding remarks in Sec. \ref{sec:conc}.

\section{System Model}\label{sec:prob}
In this section, we firstly describe our system model for the beam tracking problem and briefly summarize the Extended Kalman filter (EKF) based solution \cite{va2016beam,jayaprakasam2017robust}. 
\subsection{System Model}
We consider a scenario where an omni-directional BS is communicating with a UE that is equipped with a uniform linear array (ULA) with $N_r$ antenna elements. The ULA's response (steering vector) is given as 
\begin{equation}\label{eq:a}
a(\theta) = \frac{1}{\sqrt{N_r}}[1,e^{j\frac{2\pi}{\lambda}d \cos(\theta)},...,e^{j\frac{2\pi}{\lambda}d (N_r-1) \cos(\theta)}],
\end{equation}
where $\lambda$ is the wavelength of the carrier frequency, and $d$ is the distance between array elements. Considering a sparse system where only a single multipath component (MPC) falls into the antenna's main beam direction, the channel observation at a particular time $n$ is given as (similar to \cite{va2016beam}) 
\begin{equation}\label{eq:y}
y[n]=\frac{\alpha[n]}{N_r}\frac{1-e^{j N_r nd(\cos(\theta)-\cos(\theta_B))}}{1-e^{j nd(\cos(\theta)-\cos(\theta_B))}} + z[n],
\end{equation}
where $\alpha[n] = \alpha_R[n] + j \alpha_I[n]$ is the complex path gain, $\theta$ is the AoA of the most dominant MPC, $\theta_B$ is the direction of the main beam of the ULA, and $z[n]$ is the noise and interference from other MPC components. 

In addition to the observation model provided above, we assume that an initial connection between the BS and the UE is successfully established in the current scenario, and the fundamental problem at hand is that of maintaining this link by means of beam tracking techniques. This assumption implies the complete knowledge of initial conditions for complex gain and AoA at $n=0$. Finally, please note that the models, proposed solutions, and the results for the current paper can be easily extended to the case where BS also has ULAs, however, the current formulation was kept for simplicity. 
\subsection{Overview of EKF-based Solutions}
KF-based solutions for beam tracking systems have been explored in the literature earlier, therefore, it is important that any newly proposed solution is compared to the same. In this subsection, we briefly describe an EKF-based solution for the beam tracking problem described above that is similar in approach to \cite{va2016beam} and \cite{jayaprakasam2017robust}. To avoid implementation issues in dealing with the complex path gain the state vector for the system is formulated as
\begin{equation}
x[n]=[\alpha_R[n], ~ \alpha_I[n],~ \theta]^\top.
\end{equation}
The evolution processes for both the real and imaginary part of the path gain are assumed to be described as 
\begin{equation}
\alpha_{R/I}[n+1]=\rho\alpha_{R/I}[n]-\zeta[n],
\end{equation}
where $\rho$ is the correlation coefficient, and the $\zeta$ is a random variable distributed according to $\mathcal{N}(0,(1-\rho^2)/2)$. Since the basic recursion for EKF is fundamentally similar to the one described in \cite{va2016beam}, we do not go into its detail here, however, we note that the correlation coefficient and other statistics used for the EKF method are specifically calculated from the data generated for the current problem.
%%%%%------------------------- ML
\section{Proposed ML Solution}\label{sec:ML}
As discussed earlier, we use a recurrent NN architecture for the current beam tracking problem. The input to such a network is the observed noisy signal, $y[n]$ and the previous estimate of the AoA, $\hat{\theta}[n-1]$. The output of the network is an estimate of the AoA at the current time $\hat{\theta}[n]$.\footnote{Note that in the actual implementation we track both estimates of coefficient $\hat{\alpha}[n]$ and $\hat{\theta}[n]$, however, we report only the results of AoA tracking for clarity and due to space limitation. The elaboration and analysis for the joint tracking of both the complex path gain and the AoA is left for a future work.} While the input is restricted to $y[n]$ and $\hat{\theta}[n-1]$, the recurrent structure of the NN implicitly uses previous observations and predictions as well that make it suitable for learning over sequential data. Although the explicit input $\hat{\theta}[n-1]$ might seem redundant, it is proved to be necessary as we discuss later.

The proposed recurrent NN in this paper uses an LSTM architecture. LSTMs have been implemented successfully in several challenging tasks such as speech recognition and translation. At each time instant, the inputs to the LSTM are the observed features, the previous outputs, and the previous ``cell state". The cell state is a memory that is controlled by three gates, which control when to read, write, and erase the value of the cell. The decisions of the gates are controlled by NNs that provide nonlinear transformations of the input values.
\subsection{Custom Cost Function}
\begin{figure}[t]
\centering 
\vspace{-10mm}
\includegraphics[width=.5\textwidth]{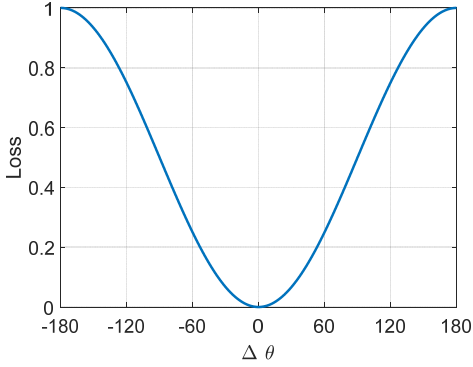}%{image} [width=0.5\textwidth]
\vspace{-10mm}
\caption{ The loss function $\mathcal{E}_\theta$ as function of $\Delta \theta$ }
\label{fig:LossFun}
\end{figure}

The weights of the above solution are determined during the training phase over a training dataset where the goal is to minimize the prediction error of the label values at the output over the observed data points. Mean Square Error (MSE) is usually used to quantify the error, where the square error $\mathcal{E}_{\rm SE}= |\hat{\theta} - \theta|^2 = |\Delta\theta| ^ 2$. However, due to the periodicity of $\theta$, large $\Delta \theta$ values might not necessary bad, for instance, $\Delta \theta = 358$ when $\theta = 1^o$ and $\hat{\theta} = 359^o$, a cost function should be able to take this into account. In this paper, we define the error function $\mathcal{E}_\theta$ as follows
\begin{equation}
    \mathcal{E}_{\rm \theta}(\Delta \theta) = \frac{1}{2}\Big(1-\cos(\Delta \theta) \Big).
\end{equation}
This cost function solves the problem of $\Delta\theta$ ambiguity that is otherwise present in MSE. For small errors, it is identical to the square error. Fig. \ref{fig:LossFun} shows a plot of this cost function. Another interesting property of the subject cost function is that it is restricted to $[0,1]$ range, where $1$ represents the highest error value. This adds more stability to the training process our error is bounded for the current problem.

\subsection{The Network}
\begin{figure}[t]
\centering 
\vspace{-9mm}
\includegraphics[width=.42\textwidth]{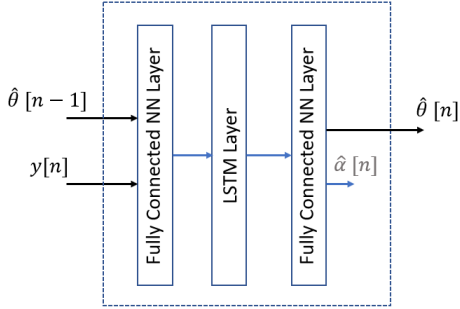}%{image} [width=0.5\textwidth]
\vspace{-10mm}
\caption{ The ML network structure. }
\label{fig:MLSolution}
\end{figure}

Fig. \ref{fig:MLSolution} shows the used ML solution, which consists of one fully connected layer with 20 neurons, one LSTM layer with 40 hidden units, followed by another fully connected layer with 20 neurons. We do not use activation functions after the fully connected layer. Further details of the input and output of the proposed network are discussed in subsection \ref{subsec:TT}.%Finally, the output layer a regression layer associated with the customized cost function.   

\section{Data Generation}\label{sec:Data}
The QuaDRiGa framework implements several channel models and allows for spatial consistency of generated sequences, which is necessary for such sequential data \cite{jaeckel2014quadriga}. Therefore, in this paper, we consider a Non-Line Of Sight (NLOS) urban micro-cell environment within this framework where the channel realizations are generated based on the mmMAGIC model \cite{mmMAGIC}, this propagation model is supported by a number of channel measurement campaigns over several frequencies \cite{mmMAGIC}. 

To generate the data for the current study, we consider the following setup. A cell has a single BS equipped with a single omnidirectional antenna located in the center of a cell with $60$ m radius and the carrier frequency is assumed to be $28$ GHz. We generate $2500$ realization of UE mobility where each of these are around $20$ m long, in a straight line, with a random initial starting location and orientation. We record the AoA and the channel coefficients of the MPCs every $0.1$ m of each path realization.

\subsection{Training and Testing}\label{subsec:TT}
For this supervised learning problem, we need to create features and labels sets before proceeding with the training and testing steps. Ideally, we would like the features $\mathcal{F}_{i,n}$ be $[{\rm real}(y_i[n]), {\rm imag}(y_i[n]), \theta_{i}[n-1]]$ and the label $\mathcal{L}_{i,n}$ be $[\theta_{i}[n]]$,\footnote{As highlight above, the actual implementation uses predicts $\alpha[n]$ (as can be seen in Fig. \ref{fig:MLSolution}), but we drop that from or discussion here.} for the training sample at time $n$ of the $i^{th}$ training sequence. However, for a beam tracking problem, the definition of the labels is challenging, as the true label depends on the beam-steering decision of the UE in a previous time-step. To see this, note that $y[n]$ depends on the true $\theta[n-1]$ by setting $\theta_B = \theta[n-1]$ in (\ref{eq:y}), however, since in reality the UE has access to $\hat{\theta}[n-1]$, the observed feature $[y[n], \hat{\theta}[n-1]]$ will be different for different ${\theta}[n-1]$. To alleviate this problem, we consider the following. We add noise to the true $\theta[n-1]$ and generate $y[n]$ accordingly. In other words, we let the training features be:
$$\mathcal{F}_{i,n} = [{\rm real}(y_{{\rm trn}, i}[n]), {\rm imag}(y_{{\rm trn}, i}[n]), \theta_{{\rm trn},i}[n-1]],$$
we here introduced the subscript $\rm ._{trn}$ to emphasize that these values are different from the one we generated above, in particular, we use $\theta_{{\rm trn},i}[n-1] = \theta_i[n-1] + \psi[n-1]$, where $\psi[n-1]\sim \mathcal{N}(0,\sigma^2_{\psi})$. We here choose $\sigma_{\psi} = 5^o $, we assume that $\psi[n-1]$ is generated independently for all $n$.

During the testing, the observed features are $[{\rm real}(y_i[n]), {\rm imag}(y_i[n]), \hat{\theta}_{i}[n-1]]$, throughout the testing phase we assume that, for a given threshold value $\theta_{\rm Th}$, when $|\hat{\theta}_{i}[n]-\theta_{i}[n]|> \theta_{\rm Th}$, the UE announces an outage and initializes a conventional AoA search/estimation procedure through a full search. In such cases, the input to the ML solution (and comparable EKF estimation) at the next time-frame will be based on the true angle once again. 

Finally, please note that we split our dataset by selecting $1750$ sequences for training and $750$ for testing and used the Adam optimizer for the network training \cite{kingma2014adam}. 
%%----------------------------------
\section{Results}\label{sec:Results}
\begin{figure}[t]
\centering 
\vspace{-35mm}

\includegraphics[width=.49\textwidth]{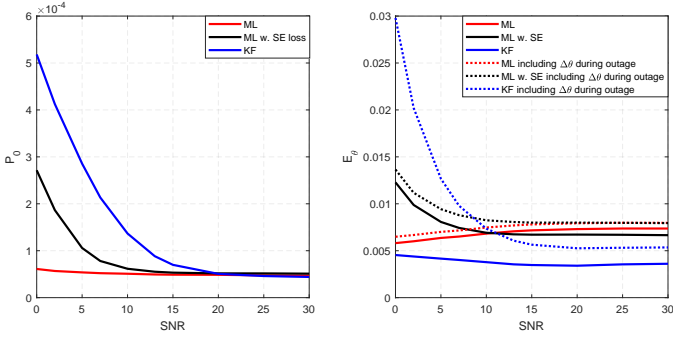}%{image} [width=0.5\textwidth]
\vspace{-35mm}
\caption{The performance vs test SNR: (a) Outage Probability $P_0$ , (b) Angle Error $\bar{E}_{\theta}$. Results are for training SNR $=7$ dB.}
\label{fig:vsSNR}
\end{figure}

In this section, we report the performance of the proposed ML solution by analyzing the outage probability and the loss values. Outage probability for the network is given as
\begin{align}
    P_0 = \frac{1}{\sum_i N_{{\rm tst},i}} \sum_i  \sum_j {\bf I}\bigg(|\Delta\theta_{i,j}| > \theta_{\rm Th} \bigg),
\end{align} where $ N_{{\rm tst},i}$ is the number of training samples for sequence $i$, ${\bf I}$ is an indicator function, and $\Delta\theta_{i,j} = \theta_{i,j}- \hat{\theta}_{i,j}$, where $\theta_{i,j}$ is the AoA at $j^{\rm th}$ instant of the $i^{\rm th}$ training sequence. We choose the threshold angle $\theta_{\rm Th}$  as a function of number of UE antennas $N_r$ as $\theta_{\rm Th} = \frac{2}{3}\frac{360}{N_r}$. The second quantity of interest is the average error, is it simply defined as
\begin{equation}
    \bar{{E}}_\theta =  \frac{1}{\sum_i N_{{\rm tst},i}} \sum_i  \sum_j \mathcal{E}_{{\rm \theta}}(\Delta \theta_{i,j}).
\end{equation}
Fig. \ref{fig:vsSNR}, shows the performance of outage probability and average error for different values of the test signal to noise ratio (SNR).  where the training is done at SNR $ = 7$ dB. We notice that the proposed ML solution shows higher robustness for smaller SNR values, while it converges to similar outage probability as EKF solution for other cases. This shows a clear advantage of the proposed ML scheme. Note that in Fig. \ref{fig:vsSNR}-(b), we plot $\bar{{E}}_\theta$ for two cases depending on whether we consider the angles during the outage. Since EKF suffers from more outages, when we exclude the angle estimates during the outage, the loss value drops significantly, however, this is not an accurate measure for the cases where the outage probability is different. For large SNR, we notice that EKF shows smaller loss value, which could indicate that EKF shows better tracking of the angles when it is not in an outage. Nevertheless, in reality, a combination of $P_0$ and $\bar{E}_{\theta}$ is important to properly analyze the performance.

\begin{figure}[t]
\centering 
\vspace{-35mm}
\includegraphics[width=.49\textwidth]{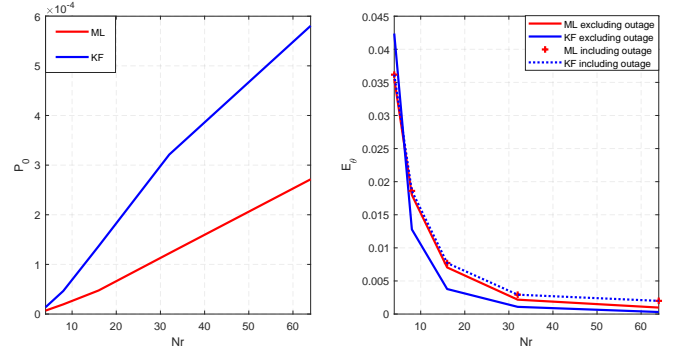}%{image} [width=0.5\textwidth]
\vspace{-35mm}
\caption{The impact of number of UE antenna $N_r$ on the performance (a) Outage Probability $P_0$ , (b) Angle Error $\bar{E}_{\theta}$. Results are for test SNR $=10$ dB.}
\label{fig:vsNr}
\end{figure}

In Fig. \ref{fig:vsNr} we consider the performance of the solution against the number of receiver antenna, $N_r$. As discussed above, $\theta_{\rm Th}$ is defined as function of $N_r$.  The testing SNR here is $10$ dB for this case. We note that the proposed ML solution shows a lower outage probability as a function of $N_r$. The values of the loss function are comparable at this SNR value, as it also can be seen from Fig. \ref{fig:vsSNR}. Note that the decay in the loss value is expected as (i) we have small beam-width (ii) frequent outages result in more angle corrections.

\begin{figure}[b]
\centering 
\vspace{-35mm}
\includegraphics[width=.49\textwidth]{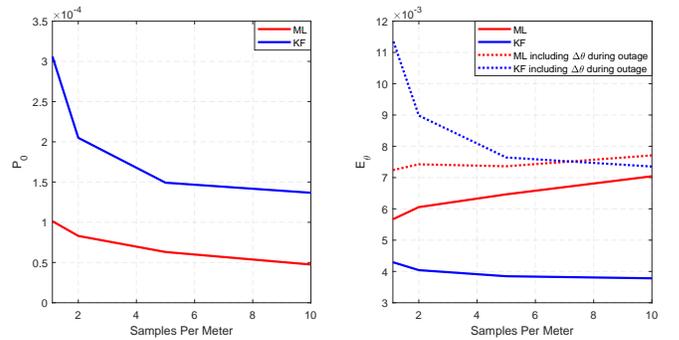}%{image} [width=0.5\textwidth]
\vspace{-35mm}
\caption{The impact of observation density on the performance (a) Outage Probability $P_0$ , (b) Angle Error $\bar{E}_{\theta}$. Results are for test SNR $=10$ dB.}
\label{fig:vsSamples}
\end{figure}

In Fig. \ref{fig:vsSamples}, we plot the performance when we vary the number of samples per meter at test and observe that the ML solution shows smaller outage probability even when smaller number of samples are used. We notice that the loss values slightly increase for ML solution as we increase the number of samples, while they decrease for the EKF solution. This could be attributed to our previous observation; the EKF shows better tracking of the AoA angle when not it is not in an outage. 

\begin{figure}[t]
\centering 
\vspace{-35 mm}
\includegraphics[width=.5\textwidth]{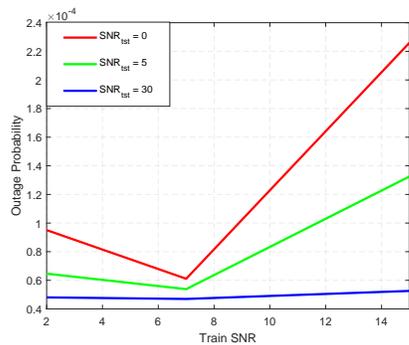}
\vspace{-40 mm}
\caption{Outage probability $P_0$ over different training SNR values for several test SNR.}
\vspace{-0 mm}
\label{fig:vsTrainSNR}

\end{figure}

Finally,  we notice that the performance of the ML solution will be impacted by the training SNR. Fig. \ref{fig:vsTrainSNR} shows $P_0$ as a function of the training SNR. We notice that moderate noise levels in training signals improve the performance, this has been noticed in the ML literature as it guards against overfitting and improves generalization. 

%%-------------------
\section{Conclusions}\label{sec:conc}
In this paper, we propose a ML-based beam tracking solution that tracks the AoA for a mobile user. The proposed solution is based a recurrent NN and two fully connected NNs. Due to the periodicity of the AoA, we propose a customized loss function based on cosine function.
The performance of the solution is studied in a standards-based stochastic environment that we generated at mmWave frequency using the QuaDRiGa framework. The results show a promising performance of the ML solution when compared to an EKF-based solution. In particular, we observe smaller outage probability, especially at low to moderate SNR.
%There are several interesting future directions, methods to reduce the loss values, impact of the loss function on the convergence,  
\pagenumbering{arabic}
\renewcommand{\thepage} {\arabic{page}}
\bibliographystyle{IEEEtran}
\bibliography{Ref}
 \end{document}